%Paper: astro-ph/9312055
%From: MOSCARDINI@ASTRPD.UNIPD.IT
%Date: Wed, 22 Dec 1993 11:01:06 +0200

% Macros for The TeXbook

\catcode`@=11 % borrow the private macros of PLAIN (with care)

\font\tentex=cmtex10

\font\ninerm=cmr9
\font\eightrm=cmr8
\font\sixrm=cmr6

\font\ninei=cmmi9
\font\eighti=cmmi8
\font\sixi=cmmi6
\skewchar\ninei='177 \skewchar\eighti='177 \skewchar\sixi='177

\font\ninesy=cmsy9
\font\eightsy=cmsy8
\font\sixsy=cmsy6
\skewchar\ninesy='60 \skewchar\eightsy='60 \skewchar\sixsy='60

\font\eightss=cmssq8

\font\eightssi=cmssqi8

\font\ninebf=cmbx9
\font\eightbf=cmbx8
\font\sixbf=cmbx6

\font\ninett=cmtt9
\font\eighttt=cmtt8

\hyphenchar\tentt=-1 % inhibit hyphenation in typewriter type
\hyphenchar\ninett=-1
\hyphenchar\eighttt=-1

\font\ninesl=cmsl9
\font\eightsl=cmsl8

\font\nineit=cmti9
\font\eightit=cmti8

\font\tenu=cmu10 % unslanted text italic
\font\magnifiedfiverm=cmr5 at 10pt

\newskip\ttglue
\def\tenpoint{\def\rm{\fam0\tenrm}%
  \textfont0=\tenrm \scriptfont0=\sevenrm \scriptscriptfont0=\fiverm
  \textfont1=\teni \scriptfont1=\seveni \scriptscriptfont1=\fivei
  \textfont2=\tensy \scriptfont2=\sevensy \scriptscriptfont2=\fivesy
  \textfont3=\tenex \scriptfont3=\tenex \scriptscriptfont3=\tenex
  \def\it{\fam\itfam\tenit}%
  \textfont\itfam=\tenit
  \def\sl{\fam\slfam\tensl}%
  \textfont\slfam=\tensl
  \def\bf{\fam\bffam\tenbf}%
  \textfont\bffam=\tenbf \scriptfont\bffam=\sevenbf
   \scriptscriptfont\bffam=\fivebf
  \def\tt{\fam\ttfam\tentt}%
  \textfont\ttfam=\tentt
  \tt \ttglue=.5em plus.25em minus.15em
  \normalbaselineskip=12pt
  \def\MF{{\manual META}\-{\manual FONT}}%
  \let\sc=\eightrm
  \let\big=\tenbig
  \normalbaselines\rm}

\def\ninepoint{\def\rm{\fam0\ninerm}%
  \textfont0=\ninerm \scriptfont0=\sixrm \scriptscriptfont0=\fiverm
  \textfont1=\ninei \scriptfont1=\sixi \scriptscriptfont1=\fivei
  \textfont2=\ninesy \scriptfont2=\sixsy \scriptscriptfont2=\fivesy
  \textfont3=\tenex \scriptfont3=\tenex \scriptscriptfont3=\tenex
  \def\it{\fam\itfam\nineit}%
  \textfont\itfam=\nineit
  \def\sl{\fam\slfam\ninesl}%
  \textfont\slfam=\ninesl
  \def\bf{\fam\bffam\ninebf}%
  \textfont\bffam=\ninebf \scriptfont\bffam=\sixbf
   \scriptscriptfont\bffam=\fivebf
  \def\tt{\fam\ttfam\ninett}%
  \textfont\ttfam=\ninett
  \tt \ttglue=.5em plus.25em minus.15em
  \normalbaselineskip=11pt
  \def\MF{{\manual hijk}\-{\manual lmnj}}%
  \let\sc=\sevenrm
  \let\big=\ninebig
  \normalbaselines\rm}

\def\eightpoint{\def\rm{\fam0\eightrm}%
  \textfont0=\eightrm \scriptfont0=\sixrm \scriptscriptfont0=\fiverm
  \textfont1=\eighti \scriptfont1=\sixi \scriptscriptfont1=\fivei
  \textfont2=\eightsy \scriptfont2=\sixsy \scriptscriptfont2=\fivesy
  \textfont3=\tenex \scriptfont3=\tenex \scriptscriptfont3=\tenex
  \def\it{\fam\itfam\eightit}%
  \textfont\itfam=\eightit
  \def\sl{\fam\slfam\eightsl}%
  \textfont\slfam=\eightsl
  \def\bf{\fam\bffam\eightbf}%
  \textfont\bffam=\eightbf \scriptfont\bffam=\sixbf
   \scriptscriptfont\bffam=\fivebf
  \def\tt{\fam\ttfam\eighttt}%
  \textfont\ttfam=\eighttt
  \tt \ttglue=.5em plus.25em minus.15em
  \normalbaselineskip=9pt
  \def\MF{{\manual opqr}\-{\manual stuq}}%
  \let\sc=\sixrm
  \let\big=\eightbig
  \normalbaselines\rm}
\magnification=1200
\hoffset 0.5truecm
\tolerance=10000
\hsize 16truecm
\vsize 24truecm
\baselineskip 0.6cm
\nopagenumbers
\font\cub=cmbx12
\def\v{\vec }
\def\ref{\par\noindent\hangindent 20pt}
\def\refig{\par\noindent\hangindent 8.5pt}
\def\mincir{\raise -
2.truept\hbox{\rlap{\hbox{$\sim$}}\raise5.truept
\hbox{$<$}\ }}
\def\magcir{\raise -
2.truept\hbox{\rlap{\hbox{$\sim$}}\raise5.truept
\hbox{$>$}\ }}
\def\gr{\kern 2pt\hbox{}^\circ{\kern -2pt K}} %  ====> GRADI KELVIN
\def\refn#1{$\null^{[#1]}$}
\def\asymp{\raise -4.3truept\hbox{$ \
\widetilde{\phantom{xy}} \ $}}
\null
\vskip 1.truecm
\centerline{\cub The Variance of QSO Counts in Cells}
\bigskip
\bigskip
\bigskip
\bigskip
\centerline{{\bf Paola Andreani}$^1$, ~{\bf Stefano Cristiani}$^1$,
{}~{\bf Francesco Lucchin}$^1$,}
\medskip
\centerline{{\bf Sabino Matarrese}$^2$ ~and ~{\bf Lauro Moscardini}$^1$}

\bigskip
\bigskip
\bigskip
\bigskip
\centerline{$^1$ {\it Dipartimento  di Astronomia, Universit\`a di Padova,}}

\centerline{{\it vicolo dell'Osservatorio 5, I--35122 Padova, Italy}}

\bigskip

\centerline{$^2$ {\it Dipartimento di Fisica ``G. Galilei", Universit\`a di
Padova,}}

\centerline{{\it via Marzolo 8, I--35131 Padova, Italy}}

\bigskip
\bigskip
\bigskip
\bigskip
\bigskip
\vfill
\centerline{Submitted to Astrophysical Journal, December 1993}
\centerline{Preprint ASTRO-PH/9312055}
\eject
\nopagenumbers
\headline={\hss\tenrm\folio\hss}
\baselineskip 0.6cm

\noindent
{\bf Abstract} \bigskip
\bigskip
\par
\noindent
{}From three quasar samples with a total of 1038 objects in the redshift range
$1.0 \div 2.2$ we measure the variance $\sigma^2$ of counts in cells of volume
$V_u$. By a maximum likelihood analysis applied separately on these samples we
obtain estimates of $\sigma^2(\ell)$, with $\ell \equiv V_u^{1/3}$. The
analysis from a single catalog for $\ell = ~40~h^{-1}$ Mpc and from a suitable
average over the three catalogs for $\ell = ~60,~80$ and $100~h^{-1}$ Mpc,
gives $\sigma^2(\ell) = 0.46^{+0.27}_{-0.27}$, $0.18^{+0.14}_{-0.15}$,
$0.05^{+0.14}_{-0.05}$ and $0.12^{+0.13}_{-0.12}$, respectively, where the
$70\%$ confidence ranges account for both sampling errors and statistical
fluctuations in the counts. This allows a comparison of QSO clustering on large
scales with analogous data recently obtained both for optical and IRAS
galaxies: QSOs seem to be more clustered than these galaxies by a biasing
factor $b_{QSO}/b_{gal} \sim 1.4 - 2.3$.
\medskip
\noindent
{\it Subject headings:} galaxies: clustering --- quasars: general,
surveys --- large--scale structure of the universe

\vfill\eject
\noindent
{\bf 1. Introduction}
\bigskip

Only in recent years the rapid growth of quasar surveys has made possible the
analysis of their clustering properties. The availability of faint quasar
samples, with their high surface density and size, has allowed a detailed study
at scales $r\leq 150~ h^{-1}$ Mpc (e.g. Shanks et al. 1987; Anderson, Kunth, \&
Sargent 1988; Iovino \& Shaver 1988; Andreani, Cristiani, \& La Franca 1991).
There is now substantial agreement on the results of the quasar two--point
correlation function $\xi(r)$. This function is larger than unity at scales $r
< 10 ~h^{-1}$ Mpc, but the issue of its evolution with redshift is still matter
of debate (Iovino, Shaver, \& Cristiani 1991; Boyle et al. 1991; Andreani \&
Cristiani 1992).

In this work we analyze QSO clustering by means of the variance of counts in
cells. The advantage of this method is to provide information on clustering at
various scales (i.e. various cell sizes), even when the volume covered by the
catalog does not form a connected region; this is particularly useful for the
available QSO samples. Statistics of counts in cells have been recently
considered by various authors (e.g. Efstathiou et al. 1990; Saunders et al.
1991; Loveday et al. 1992; Gazta\~naga 1992; Bouchet et al. 1993), to obtain
reliable constraints on the amplitude of galaxy clustering on different scales,
through the variance, and on its possible deviations from a Gaussian behavior,
through higher order moments such as the skewness. On the other hand, it is
relatively easier, within a model for structure formation, to obtain
theoretical predictions for the moments of counts in cells at various scales.
Moreover, this kind of analysis, combined with similar studies performed for
optical and IRAS galaxies, allows a direct determination of the biasing factor
relating the clustering of QSOs with that of these classes of objects.

After shot--noise subtraction, the variance of the continuous density
fluctuation field, smoothed over the cell size $\ell$, is related to the
spatial correlation function $\xi(r)$ by the integral
$$
\sigma^2(\ell) = \int_0^\infty d r~r^2 \xi(r) ~{\cal F}_\ell(r),
\eqno(1)
$$
where the window function ${\cal F}_\ell(r)$ takes into account the details of
the cell geometry. For spherical cells of radius $R$, one finds
$$
{\cal F}_R(r) = {18 \over \pi R^3} \int_0^\infty dx~j_1^2(x) ~j_0(xr/R)
\approx {3 \over R^3} \vartheta_H(R-r),
\eqno(2)
$$
where $j_\ell$ are spherical Bessel functions of order $\ell$ and
$\vartheta_H(x)$ is the Heaviside function (which is zero for $x<0$ and one for
$x>0$). These relations allow to connect the results of this work with previous
data on the quasar--quasar correlation function. Actually, the two methods are
complementary: the variance yields a more compact information on the clustering
amplitude at the scale of the cell--size, while the correlation function gives
a more detailed geometrical information. Being a volume average of the
correlation function, the variance is characterized by a higher
signal--to--noise ratio.

\bigskip
\noindent
{\bf 2. Data Samples and Statistical Analysis}
\medskip

Table I lists our database, which consists of eight different surveys already
published. Table I reports the sample name (column 1), the effective covered
area (column 2), the limiting magnitude (column 3), the number of objects with
$M_B \le -23$ \footnote {$^\dagger$} {\eightpoint The absolute $B$ magnitude is
calculated assuming Hubble constant $H_0 = 100~h$ km s$^{-1}$ Mpc$^{-1}$, with
$h=0.5$ in a flat universe, with vanishing cosmological constant;
$k$--corrections are as in Cristiani \& Vio (1990) and galactic extinction as
in Burstein \& Heiles (1982).} (column 4), within the assumed redshift range
(column 5), and the number of objects between redshift $1$ and $2.2$ (column
6).

The samples contain objects selected with different techniques: UV--excess,
variability and slitless spectroscopy. Attention has been paid to use only
complete catalogues, in order to minimize systematic biases. The optimal
redshift range for our statistical study is $1 - 2.2$: this is because the
highest QSO number density is in this redshift range and the catalog
completeness decreases beyond $z=2.2$.

In spite of the different catalog selection criteria, the high completeness in
the considered redshift range allows to subdivide our database in three groups
(named Sample A, B and C, in the following) on the only basis of their limiting
magnitude; each of these samples will then be characterized by its own
selection function. Sample A ($510$ objects): APM; Sample B ($332$ objects):
Boyle et al. (1990), $m_J \leq 21$ sample (hereafter HVI) from Hawkins \&
V\'eron (1993), Zitelli et al. (1992) and Osmer \& Hewett (1991), all cut at
the limiting magnitude $m_J=20.85$, which leads to a $2.5\%$ decrease in the
number of objects; Sample C ($122$ objects): La Franca et al. (1992), $m_J \leq
19.5$ sample (hereafter HVII) from Hawkins \& V\'eron (1993) and Crampton et
al. (1989), cut at $m_J=19.5$, with a $34\%$ decrease in the number. The actual
limiting magnitude has been chosen slightly different for each sample, to take
into account the different galactic extinction. The $B$ magnitudes of La Franca
et al. (1992) have been converted to $J$ magnitudes, according to the relation
$m_J=m_B-0.05$.

To compute the moments of QSO counts in cells we first divide our three samples
in shells with mean radii $r_a$ centered on the observer, further divided in
$M_a$ cells of volume $V_u$. Let $N_j$ be the number of objects in the $j$--th
cell ($j=1, \dots, M_a$) of a given shell and $V_j \leq V_u$ the cell volume
actually included in the sample boundary, estimated by means of a standard
Monte Carlo technique. Cells with $V_j < 0.5 V_u$ have not been used.

In calculating the variance of counts in cells we had to account for the volume
incompleteness of our samples. Following Efstathiou et al. (1990) we write
$$
\sigma^2_a \equiv \overline \Sigma_a^2 =
{\sum_j (N_j - V_j {\overline N}_a / V_u)^2 - (1 - \sum_k V_k^2/(\sum_k V_k)^2)
\sum_j N_j
\over ({\overline N}_a / V_u)^2~[\sum_k V_k^2
- 2 \sum_k V_k^3 / \sum_k V_k + (\sum_k V_k^2)^2/
(\sum_k V_k)^2]},
\eqno(3)
$$
where $\Sigma^2_a = (\Delta N/\overline N_a)^2$ and ${\overline N}_a = V_u
\sum_j N_j / \sum_j V_j$ is the expected number of objects in a cell of volume
$V_u$ belonging to the $a$--th shell. The shot--noise subtraction in Eq.(3) may
result in negative values for the estimates of $\sigma^2_a$ (see, e.g., Figure
1 in Efstathiou et al. 1990): this is because the Poisson model only
approximately describes discreteness effects. In this sense one can safely
state that Eq.(3) represents an estimate of the excess variance above the
Poisson level; this also assumes that the expected variance is independent of
the cell--volume, which is the case if the missing volume in incomplete cells
is small, given that $\sigma^2_a$ is likely to be a weak function of cell size.

When using different catalogs grouped within a sample (as in Samples B and C),
even if the selection method is the same, the effects of systematic errors
(e.g. in the zero point of the magnitude calibration) have to be considered. We
have therefore normalized the different catalogs of Samples B and C by
selecting as a reference catalog the one with the highest surface density in
the sample and reducing the effective cell volumes of the remaining ones by the
ratio of their surface density (derived from Table I) to that of the reference
catalog. In this way, we expect to have removed the above mentioned systematic
effects, leaving a bias, if any, in the sense of underestimating the variance.

Errors in the estimate of the variance, ${\rm Var}(\Sigma^2)$, are computed by
the quadratic sum of two terms: a first one, ${\rm Var}_{samp}(\Sigma^2)$,
accounting for the sampling errors inherent in our data, and a second one,
${\rm Var}_{stat}(\Sigma^2)$, corresponding to the statistical uncertainty. In
order to quantify the sampling errors in our data we used a bootstrap
resampling technique (e.g. Barrow, Bhavsar, \& Sonoda 1984) in each separate
sub--sample, accounting for the different densities. The second contribution to
the variance of $\Sigma^2_a$ can be estimated under the simplifying assumptions
that the cells are independent; making then use of the Central Limit Theorem
one can approximate the underlying distribution by a Gaussian with variance
$\sigma^2_a = \overline \Sigma_a^2$ (see Efstathiou et al. 1990). This results
in
$$
{\rm Var}_{stat}(\Sigma^2_a) = {2~(1 + \sigma^2_a) + 4~{\overline
N}_a \sigma^2_a + 2~({\overline N}_a)^2 \sigma^4_a \over M_a
({\overline N}_a)^2 }.
\eqno(4)
$$
This method of calculating ${\rm Var}(\Sigma^2)$ allows to deal with catalogs
characterized by both reduced number of cells (such as Sample C) and dilution
effects (Sample A): in the former case the larger contribution comes from the
theoretical variance, in the latter one from the bootstrap error. Note,
however, that this method leads to a more conservative estimate of error bars,
which result typically higher than in previous analyses of the variance of
counts in cells.

The final variance, $\sigma^2$, for the cell counts of QSOs at a given scale,
separately for samples A, B and C, is obtained by  maximizing the likelihood
function:
$$
{\cal L} (\sigma^2) = \prod_a {1 \over [2\pi
{\rm Var}(\Sigma^2_a)]^{1/2}}
\exp\biggl[-{(\sigma^2_a - \sigma^2)^2 \over 2~{\rm Var}(\Sigma^2_a)}\biggr],
\eqno(5)
$$
where the product extends over all shells.

\bigskip
\noindent
{\bf 3. Results and Discussion}
\medskip

We report the results for the variance of counts in cells of sizes $\ell \equiv
V_u^{1/3} = 60,~80$ and $100~h^{-1}$ Mpc. For sample B, which is the one of
highest density, we can also consider cells of size $\ell = 40~h^{-1}$ Mpc. All
these cells are obtained with parallelepiped--shaped geometry, with
line--of--sight dimension larger than the transversal ones by a factor of
$1.55$, in order to better follow the geometry of the catalogs.

Figures 1, 2 and 3 show, for the considered cell--sizes, the variance of QSO
counts in cells for Sample A, B and C respectively, obtained from Eq.(3), with
error bars given by ${\rm Var}(\Sigma^2)$. The maximum likelihood estimates of
the variance as a function of the cell--size for the three samples separately
are reported in Table II; the $70\%$ errors are obtained by computing the
values of the variance where the likelihood in Eq.(5) drops by a factor of
$1.71$ from its maximum. When the lowest value becomes negative we consistently
replace it with zero. Table II also shows the $\chi^2$ values and the number of
radial shells $N_s$ for each determination.

Having consistently computed three independent estimates of $\sigma^2$ at
various scales, the maximum likelihood method [Eq.(5)] can now be used to
estimate the overall variance, considering all samples together, at $60$, $80$
and $100~h^{-1}$ Mpc: we find $\sigma^2 = 0.18^{+0.14}_{-0.15}$,
$0.05^{+0.14}_{-0.05}$ and $0.12^{+0.13}_{-0.12}$, ($70\%$ confidence range),
respectively. We can also compare these data with the estimate of the variance
resulting from the QSO correlation function, $\xi(r)$, obtained from Sample A,
B and C, separately, according to the methods described in Andreani, Cristiani,
\& La Franca (1991). We fit a spline to $\xi(r)$ and numerically integrate
Eq.(1) and Eq.(2), with a spherical top--hat filter of equivalent volume. The
results and the errors, obtained by bootstrap resampling, are shown in Figure 4
together with the maximum likelihood estimates of $\sigma^2$ from the counts.
Within the ($70\%$) error bars, the two methods provide compatible results,
although the values of $\sigma^2$ derived from the counts in Sample B are
systematically higher.

In order to evaluate the possible redshift dependence of the clustering
amplitude we calculated the variance for the two separated redshift ranges $1
\leq z \leq 1.6$ and $1.6 \leq z \leq 2.2$. We found that the results are
compatible with a constant comoving clustering amplitude within the error bars
(e.g. Andreani \& Cristiani 1992), although a slight tendency to have larger
variance in the nearest strip occurs in Sample B and C.

Given the size of the error bars, which in some cases make the results
compatible with no clustering, we decided to check the presence of real
clustering in our data by performing a Kolmogorov--Smirnov test against the
null hypothesis that our counts are drawn from a Poisson parent distribution.
To this aim we generated $10,000$ mock Poisson catalogs with the same density,
selection function and volume coverage, separately for Sample A, B, and C. We
then compared the resulting histograms of the cell counts with the real ones.
They are shown in Figure 5, where the dotted lines correspond to histograms of
the counts in cell for the Poissonian case and the solid lines to those of the
real data for the three samples and different cell sizes. The
Kolmogorov--Smirnov test indicates that the Poisson hypothesis can be rejected
at a high confidence level ($10^{-20}$ at $40~h^{-1}$ Mpc and down to $10^{-2}$
at $100~h^{-1}$ Mpc), with the only exception of Sample A, for which the
Poissonian hypothesis cannot be rejected; in this case indeed we found that the
bootstrap errors dominate the overall variance of $\Sigma^2$.

Our results can be compared with those for the variance of IRAS galaxies in the
QDOT sample, analyzed by Efstathiou et al. (1990); they find $\sigma^2 =
0.21^{+0.11}_{-0.07}$ and $0.05^{+0.06}_{-0.03}$, for cubic cells of size $40$
and $60~h^{-1}$ Mpc respectively. Loveday et al. (1992), who performed a
similar analysis in the Stromlo--APM redshift survey of optical galaxies,
obtain $\sigma^2 = 0.14^{+0.08}_{-0.05}$, $0.05^{+0.06}_{-0.03}$ and
$0.02^{+0.05}_{-0.01}$, for cubic cells of size $40$, $60$ and $75~h^{-1}$ Mpc,
respectively. A recent estimate is given by Bouchet et al. (1993) for the $1.2$
Jy IRAS Galaxy Redshift Survey; they get the best--fit $\log \sigma^2(R) =
(1.17 \pm 0.05) - (1.59 \pm 0.06) \log R$, for spherical cells of radius $R$.
This corresponds to $\sigma^2 \approx 0.09$ and $0.05$, for $\ell=40$ and
$60~h^{-1}$ Mpc respectively (having accounted for the different geometry of
the cells). Note that the ($95\%$) confidence ranges quoted by Efstathiou et
al. (1990) and Loveday et al. (1992) are obtained by considering only the
theoretical part of the error, i.e. neglecting sampling fluctuations, whilst we
made the more conservative choice of summing up the two uncertainties.

Our data are compatible, within the errors, with all results above.
Nevertheless, it could be argued that QSOs are biased over both IRAS and
optical galaxies; we find $b_{QSO}/b_{gal}=\sigma_{QSO}/\sigma_{gal}$ in the
range $1.4 - 2.3$. This effect is indeed predicted by hierarchical theories of
quasar formation within massive haloes (Efstathiou \& Rees 1988; Cole \& Kaiser
1989; Haehnelt \& Rees 1993), although the amplitude of such a bias strongly
depends on the specific model of structure formation. This issue clearly
deserves more realistic modeling of quasar origin, also taking into account the
recent observational constraints from large--scale structures, such as the
normalization implied by COBE data (Smoot et al. 1992). On the other hand, our
statistical analysis shows that more stringent constraints on quasar clustering
will only be obtained when new catalogs will be constructed with homogeneous
selection criteria and over wider and deeper regions of the sky: a goal which
can be reached within few years, thanks to the availability of multiobject
spectrographs.

\bigskip
\noindent
{\bf Acknowledgments}
\medskip
Will Saunders and Roberto Vio are gratefully acknowledged for helpful
suggestions and discussions. The anonymous referee is also thanked for useful
comments. We acknowledge the Ministero Italiano dell'Universit\`{a} e della
Ricerca Scientifica e Tecnologica for financial support. P.A. also thanks
partial support by CNR, under contract $92.01346.{\rm CT}.02$.
\vfill\eject

\centerline{\bf REFERENCES}
\medskip

\ref {Anderson, N., Kunth, D., \& Sargent, W.L.W. 1988,  AJ, 95, 644}

\ref {Andreani, P., Cristiani, S., \&  La Franca, F. 1991, MNRAS, 253, 527}

\ref {Andreani, P., \& Cristiani, S. 1992, ApJ, 398, L13}

\ref {Barrow, J.D., Bhavsar, S.P., \& Sonoda, D.H. 1984,
MNRAS, 210, 19p}

\ref {Boyle, B.J., Fong, R., Shanks, T., \& Peterson, B.A. 1990, MNRAS,
243, 1}

\ref {Boyle, B.J., Jones, L.R., Shanks, T., Marano, B., Zitelli, V., \&
Zamorani,  G. 1991, in Proc. of the Meeting on The Space Distribution of
Quasars, ed. D. Crampton, ASP Conference Series, 21, p.191}

\ref {Bouchet, F.R., Strauss, M.A., Davis, M., Fisher, K.B., Yahil, A.,
\& Huchra, J.P. 1993, ApJ, 417, 36}

\ref {Burstein, D., \& Heiles, C. 1982, AJ,  87, 1165}

\ref {Chaffee, F.H., Foltz, C.B., Hewett, P.C., Francis, P.J., Weymann, R.J.,
Morris, S.L., Anderson, S.F., \& MacAlpine, G.M. 1991, AJ, 102, 461}

\ref {Cole, S., \& Kaiser, N. 1989, MNRAS, 237, 1127}

\ref {Crampton, D., Cowley, A.P., \& Hartwick, F.D.A. 1989, AJ, 345, 59}

\ref {Cristiani, S., \& Vio, R. 1990, A\&A, 227, 385}

\ref {Efstathiou, G., Kaiser, N., Saunders, W., Lawrence, A., Rowan--Robinson,
M., Ellis, R.S., \& Frenk, C.S. 1990, MNRAS, 247, 10p}

\ref {Efstathiou, G., \& Rees, M.J. 1988, MNRAS, 230, 5p}

\ref {Foltz, C.B., Chaffee, F.H., Hewett, P.C., MacAlpine, G.M., Turnshek,
D.A.,
Weymann, R.J., \& Anderson, S.F. 1987, AJ, 94, 1423}

\ref {Foltz, C.B., Chaffee, F.H., Hewett, P.C., Weymann, R.J., Anderson, S.F.,
\& MacAlpine, G.M. 1989, AJ, 98, 1959}

\ref {Gazta\~naga, E. 1992, ApJ, 398, L17}

\ref {Hawkins, M.R.S., \& V\'eron, P. 1993, MNRAS, 260, 202}

\ref {Haehnelt, M., \& Rees, M.J. 1993, preprint}

\ref {Hewett, P.C., Foltz, C.B., Chaffee, F.H., Francis, P.J., Weymann, R.J.,
Morris, S.L., Anderson, S.F., \& MacAlpine, G.M. 1991, AJ, 101, 1121}

\ref {Iovino, A., \& Shaver, P. 1988, ApJ, 330, L13}

\ref {Iovino, A., Shaver, P., \& Cristiani, S. 1991,
in Proc. of the Meeting on The Space Distribution of Quasars,
ed. D. Crampton, ASP Conference Series, 21, p.202}

\ref {La Franca, F., Cristiani, S., \& Barbieri, C. 1992,
AJ, 103, 1062}

\ref {Loveday, J., Efstathiou, G., Peterson, B.A., \& Maddox, S.J. 1992,
ApJ, 400, L43}

\ref {Morris, S.L., Weymann, R.J., Anderson, S.F., Hewett, P.C., Foltz, C.B.,
Chaffee, F.H., Francis, P.J., \& MacAlpine, G.M. 1991, AJ, 102, 1627}

\ref {Osmer, P.S., \& Hewett, P.C. 1991, ApJS, 75, 273}

\ref {Saunders, W., Frenk, C.S., Rowan--Robinson, M., Efstathiou, G.,
Lawrence, A., Kaiser, N., Ellis, R.S., Crawford, J., Xia, X.--Y.,
\& Parry, I. 1991, Nature, 349, 32}

\ref {Shanks, T., Fong, R., Boyle, B.J., \& Peterson, B.A. 1987,
MNRAS, 227, 739}

\ref {Smoot, G.F., et al. 1992, ApJ, 396, L1}

\ref {Zitelli, V., Mignoli, M., Zamorani, G., Marano, B., \& Boyle, B.J. 1992,
MNRAS, 256, 349}

\vfill\eject

\baselineskip 0.4cm
\midinsert
\vskip 1truecm
\endinsert
\hoffset -0.3truecm
\centerline  {\bf Table I}
\bigskip
\centerline {\bf Quasar surveys}
\bigskip
\settabs\+&HHHHHHHHHHHHHHHHH &HHHHHHH &HHHHHHHH &HHHHHHH &HHHHHH & HHHHHHHH
\cr
\bigskip
\hrule
\bigskip
\+& survey & surface$^\star$ & limiting & \# objects$^\dagger$ & $z$ range
& \# object in \cr
\par
\+& & sq. deg. & magnitude & & & $1\leq z \leq 2.2$ \cr
\medskip
\hrule
\medskip
\par
\+& APM$^\bullet$ & ~516 & $m_J \leq$ 18.5 & ~~~1006 & 0.2--3.1 & ~~~~~~510\cr
\medskip
\+& Boyle et al. (1990) & ~10.15$^\ddagger$& $m_J
\leq$ 20.9 & ~~~320 & 0.2--2.2 & ~~~~~~236 \cr
\medskip
\+& Crampton et al. (1989) & ~4.8 & $m_J \leq$ 20.5 & ~~~135 & 0.2--3.1 &
{}~~~~~~87\cr
\medskip
\+& HVI$^\diamond$ & ~2 & $m_J \leq$ 21.0 & ~~~29 & 0.3--2.2 &
{}~~~~~~24\cr
\par
\+& HVII$^\diamond$ & ~19 & $m_J \leq$ 19.5 & ~~~66 & 0.3--2.2 &
{}~~~~~~40\cr
\medskip
\+& La Franca et al. (1992) & ~10 & $m_B \leq$ 19.9 & ~~~95 & 0.35--2.2 &
{}~~~~~~63\cr
\medskip
\+& Osmer \& Hewett (1991) & ~6.1 & $m_J \leq$ 21.7 & ~~~113 & 0.2--3.1 &
{}~~~~~~66\cr
\medskip
\+& Zitelli et al. (1992) & ~0.69 & $m_J \leq$ 20.85 & ~~~21 & 0.6--2.8 &
{}~~~~~~12\cr
\medskip
\hrule
\medskip
\medskip
\par\noindent
{$^\star$ claimed effective area
\par\noindent
$^\dagger$ number of objects with $M_B \leq -23$ in a $h=0.5$
and $\Omega_0=1$ universe.
\par\noindent
\refig {$^\bullet$ Foltz et al. (1987); Foltz et al. (1989);
Hewett et al. (1991); Chaffee et al. (1991);
Morris et al. (1991)}
\par\noindent
$^\ddagger$ only 5 out of 8 fields have been used in this work
\par\noindent
$^\diamond$ Hawkins \& V\'eron (1993)}
\vfill\eject

\baselineskip=15truept
\centerline{\bf Table II}
\bigskip
\centerline{\bf Variance from counts in cells}
\bigskip
{\settabs 5 \columns
\+  ~~~~~~~Sample & $\ell~(h^{-1}$ Mpc)~~ & ~~$\sigma^2$  &
{}~~$\sigma^2~(70\%)$ & ~$\chi^2/N_s$ \cr
\+         &        &             &  &                 \cr
\+ ~~~~~~~~~~$A$ & $~~~~~~~60$ & $0.22$ & $0.00-0.62$ & 0.88/11 \cr
\+ & $~~~~~~~80$ & $0.00$ & $0.00-0.32$ & 0.32/8 \cr
\+ & $~~~~~~100$ & $0.00$ & $0.00-0.24$ & 0.64/6 \cr
\medskip
\+ ~~~~~~~~~~$B$ & $~~~~~~~40$ & $0.46$ & $0.19-0.73$ & 6.90/16 \cr
\+ & $~~~~~~~60$ & $0.27$ & $0.07-0.47$ & 2.72/11 \cr
\+ & $~~~~~~~80$ & $0.13$ & $0.00-0.31$ & 2.10/8 \cr
\+ & $~~~~~~100$ & $0.20$ & $0.00-0.40$ & 2.09/6 \cr
\medskip
\+ ~~~~~~~~~~$C$ & $~~~~~~~60$ & $0.00$ & $0.00-0.25$ & 4.39/11 \cr
\+ & $~~~~~~~80$ & $0.00$ & $0.00-0.22$ & 1.09/8 \cr
\+ & $~~~~~~100$ & $0.10$ & $0.00-0.37$ & 0.38/6 \cr}
\vfill\eject

\baselineskip 0.6cm
\noindent
{\bf Figure captions}
\medskip
\noindent
{\bf Figure 1}. The variance $\sigma^2$ in cells of size $\ell = 60,~80$ and
$100~h^{-1}$ Mpc as a function of redshift, for sample A. The error bars are
one standard deviation, ${\rm Var}(\Sigma^2)$. The solid line represents the
maximum likelihood estimate of the variance; the dotted lines correspond to the
$70\%$ confidence range.
\medskip
\noindent
{\bf Figure 2}. The variance $\sigma^2$ in cells of size $\ell = 40,~60,~80$
and $100~h^{-1}$ Mpc as a function of redshift, for sample B. Error bars and
lines as in Figure 1.
\medskip
\noindent
{\bf Figure 3}. The variance $\sigma^2$ in cells of size $\ell = 60,~80$ and
$100~h^{-1}$ Mpc as a function of redshift, for sample C. Error bars and lines
as in Figure 1.
\medskip
\noindent
{\bf Figure 4}. Comparison of two estimates of the QSO variance $\sigma^2$ for
the three samples: filled squares refer to the estimate obtained from the
counts in cells, open squares to the integral of the correlation function.
Error bars give the $70\%$ confidence range. For clarity, the two different
estimates are shown with a small horizontal shift.
\medskip
\noindent
{\bf Figure 5}. Histograms of the counts in cells for the three separated
samples A, B and C and for different cell sizes (from 40 to $100~h^{-1}$ Mpc).
The dotted lines correspond to the Poissonian case, while the solid ones to the
real data. The hypothesis that the distribution of objects in cells is
compatible with a Poissonian one can be rejected at a very high confidence
level for samples B and C, but not for sample A.
\medskip
\bye